\documentclass{icrc}

\usepackage{times}
\usepackage{graphicx} % when using Latex and dvips
\def\apj{{\it Astrophys. J.\ }}

\def\aap{{\it Astronomy and Astrophys.\ }}

\def\mnras{{\it Mon. Not. R. Astron. Soc.\ }}

\def\mathcal{{\it }}

\def\la{\mathrel{\mathchoice {\vcenter{\offinterlineskip\halign{\hfil
\(\displaystyle##\)\hfil\cr<\cr\sim\cr}}}
{\vcenter{\offinterlineskip\halign{\hfil\(\textstyle##\)\hfil\cr
<\cr\sim\cr}}}
{\vcenter{\offinterlineskip\halign{\hfil\(\scriptstyle##\)\hfil\cr
<\cr\sim\cr}}}
{\vcenter{\offinterlineskip\halign{\hfil\(\scriptscriptstyle##\)\hfil\cr
<\cr\sim\cr}}}}}

\def\etal{{\it et al.,\ }} % macros exists
\def\eg{{ e.g.,\ }}
\def\ie{{ i.e.,\ }}

\def\p1{{Part~I\ }}
\def\jpg{{Journ. Phys. G }}

\def\RPP{{Rep. Progr. Phys }}

\def\ln{{\rm ln}}
\def\const{{\rm const\ }}

\begin{document}

\title{Intrinsic energy cut-off in diffusive shock acceleration: \\
possible reason for non-detection
of TeV-protons in SNRs}

\author[1]{M.A. Malkov}
\author[1]{P.H. Diamond}
\affil[1]{University of California at San Diego,
9500 Gilman Dr, La Jolla, CA 92093-0319, USA}

\author[2]{T.W. Jones}
\affil[2]{University of Minnesota, 116 Church St SE Minneapolis, MN 55455}

\correspondence{M.A. Malkov (mmalkov@ucsd.edu)}

\firstpage{1}
\pubyear{2001}

% \titleheight{11cm} % uncomment and adjust in case your title block
                     % does not fit into the default and minimum 7.5 cm

\maketitle

\begin{abstract}
The linear theory of shock acceleration predicts the maximum particle
energy to be limited only by the acceleration time and the size of the
shock. We study the \emph{combined} effect of acceleration
nonlinearity (shock modification by accelerated particles, that
\emph{must} be present in strong astrophysical shocks) and propagation
of Alfven waves that are responsible for particle confinement to the
shock front. We show that wave refraction to larger wave numbers in
the nonlinearly modified flow causes enhanced losses of particles in
the momentum range
\( p_{max}/R<p<p_{max} \), where \( R>1 \) is the nonlinear 
pre-compression of the flow and
\( p_{max} \) is a conventional maximum momentum, that could 
be reached if there was no refraction.
\end{abstract}

\section{Introduction}

One of the most important parameters of the Fermi acceleration is the
rate at which it operates.  Indeed, what is often predicted or even
observed is a power-law spectrum that cuts off due to the finite
acceleration time. The cut-off momentum \( p_{max}(t) \) advances with
time as follows

\begin{equation}
\label{p:max}
\frac{dp_{max}}{dt}=\frac{p_{max}}{t_{acc}}
\end{equation}
 while the acceleration time is determined by (e.g., Axford, 1981)

\begin{equation}
\label{t:acc}
t_{acc}=\frac{3}{u_{1}-u_{2}}\int _{p_{min}}^{p_{max}}\left[
\frac{\kappa _{1}(p)}{u_{1}}+\frac{\kappa _{2}(p)}{u_{2}}\right]
\frac{dp}{p}
\end{equation}
with \( u_{1} \) and \( u_{2} \) being the upstream and downstream
flow speeds in the shock frame whereas \( \kappa _{1} \) and \( \kappa
_{2} \) are the particle diffusivities in the respective media. These
are the most sensitive quantities here which are determined by the
rate at which particles are pitch angle scattered by the Alfven
turbulence. If the latter was just  background turbulence in the
interstellar medium, the acceleration process would be too slow to
account for the galactic cosmic rays (CRs). However it was realized
(e.g., Bell, 1978) that accelerated particles must create the
scattering environment by themselves generating Alfven waves via the
cyclotron instability. This wave generation process proved to be 
very efficient
(e.g., V\"olk
\etal 1984) so that the normalized wave energy 
density \( \left( \delta B/B_{0}\right) ^{2} \)
is related to the partial pressure \( P_{c} \) of CRs that 
resonantly drive these waves through

\begin{equation}
\label{delB}
\left( \delta B/B_{0}\right) ^{2}\sim M_{A}P_{c}/\rho u^{2}
\end{equation}
where \( M_{A}\gg 1 \) is the Alfven Mach number and \( \rho u^{2} \)
is the shock ram pressure.  Often it is assumed that the turbulence
saturates at \( \delta B/B_{0}\sim 1 \), which means that the
m.f.p. of pitch angle scattered particles is of the order of their
gyro-radius \( r_{g} \).  Then, \( \kappa =\kappa _{B}\equiv c \)\(
r_{g}(p)/3 \), where \( \kappa _{B} \) stands for the Bohm diffusion
coefficient. Hence, \( t_{acc}\sim \left( eB/p\right) ^{-1}\left(
c/u_{1}\right) ^{2} \).

However the acceleration rate (\ref{t:acc}) with \( \kappa =\kappa
_{B} \) was found to be fast enough to explain the acceleration of CRs
in SNRs up to the ``knee'' energy \( \sim 10^{15}eV \) over their life
time. The analyses of Drury \etal (1994) and Naito and Takahara (1994)
of prospective detection of super-TeV emission from nearby SNR (they
must result from the decays of \( \pi ^{0} \) mesons born in
collisions of shock accelerated CR protons with the nuclei of
interstellar gas) look equally optimistic. The expected fluxes were
shown to be strong enough to be detected by the imaging Cherenkov
telescopes. Moreover, the EGRET (Esposito \etal 1996) detected a lower
energy (\( \la \Gamma eV \)) emission coinciding with some galactic
SNRs. Unfortunately, despite the physical robustness of the
above-mentioned predictions of emission, no statistically significant
signal that could be attributed to any of the EGRET sources was
detected (Buckley \etal 1997).  The further complication is that the
region between GeV and TeV energy bands is currently uncovered by any
instrument. Therefore, based on these observational results it was
suggested (e.g., Buckley
\etal 1997) that there is probably a spectral break or even cutoff somewhere within 
this band.

In this paper we attempt to understand what may happen to the spectrum
provided that the acceleration is indeed fast enough to access the TeV
energies over the life time of SNRs in question. Our starting point is
that the fast acceleration also means that the pressure of accelerated
particles becomes significant relatively early and must change the
entire shock structure by this time.  At the first glance this should
not slow down acceleration since according to eq.(\ref{delB}) this
increases the turbulence \emph{level} improving thus particle
confinement near the shock front and making acceleration
faster. Simultaneously with that but more importantly, the upstream
flow is decelerated by the pressure of CRs \( P_{c} \) which
influences the \emph{spectral properties} of the turbulence by
affecting the propagation and excitation of the Alfven waves.  This
effect is twofold. First the waves are compressed in the converging
plasma flow upstream and are thus blue-shifted lacking the long waves
needed to keep the high energy particles diffusively bound to the
accelerator. Second, and as a result of the first, at highest energies
there remain relatively few particles so that the level of resonant
waves is also small and hence the acceleration rate is low.

\section{Basic Equations and Approximations}

We use the standard diffusion-convection equation for describing the
transport of high energy particles (CRs) near a CR modified shock

\begin{equation}
\label{dc1}
\frac{\partial f}{\partial t}+U\frac{\partial f}
{\partial x}-\frac{\partial }{\partial x}\kappa 
\frac{\partial f}{\partial x}=\frac{1}{3}
\frac{\partial U}{\partial x}p\frac{\partial f}{\partial p}
\end{equation}
Here \( x \) is directed along the shock normal (also the direction of
the ambient magnetic field). The two quantities that control the
acceleration process are the flow profile \( U(x) \) and the particle
diffusivity \( \kappa (x,p) \). The first one is coupled to the
particle distribution \( f \) through the equations of mass and
momentum conservation

\begin{eqnarray}
\frac{\partial }{\partial t}\rho +\frac{\partial }{\partial x}\rho 
U=0 &  & \label{mas:c} \\
\frac{\partial }{\partial t}\rho U+\frac{\partial }{\partial x}
\left( \rho U^{2}+P_{\rm c}+P_{\rm g}\right) =0 &  & \label{mom:c} 
\end{eqnarray}
 where

\begin{equation}
\label{P_c}
P_{\rm {c}}(x)=\frac{4\pi }{3}mc^{2}\int _{p_{0}}^{\infty
}\frac{p^{4}dp}{\sqrt{p^{2}+1}}f(p,x)
\end{equation}
is the pressure of the CR gas, and \( P_{\rm g} \) is the gas
pressure. The lower boundary
\( p_{0} \) in the momentum space separates CRs from 
the thermal plasma and enters the equations
through the magnitude of \( f \) at \( p=p_{0} \) which specifies the
injection rate. The particle momentum \( p \) is normalized to \( mc
\). We assume that the upstream region is
\( x>0 \) half-space, and represent the velocity profile 
in the shock frame as \( U(x)=-u(x) \) where the (positive) flow speed
\( u(x) \) jumps from \( u_{2}\equiv u(0-) \) downstream to \(
u_{0}\equiv u(0+)>u_{2} \) across the subshock and then gradually
increases up to \( u_{1}\equiv u(+\infty )\geq u_{0} \) (see Fig.1a).
Limiting our consideration to high Mach number shocks, \( M\gg 1 \),
we may drop \( P_{\rm g} \) term \emph{upstream, \( x>0 \)}. It is
retained at the subshock which, however, can be described by the
conventional Rankine-Hugoniot jump condition

\begin{equation}
\label{c:r}
\frac{u_{0}}{u_{2}}=\frac{\gamma +1}{\gamma -1+2M_{0}^{-2}}
\end{equation}
 where \( M_{0} \) is the Mach number in front of the subshock. For
 simplicity we use the adiabatic approximation, \ie the far upstream
 Mach number is related to \( M_{0} \) by \( M_{0}^{2}=M^{2}/R^{\gamma
 +1} \), where \( R\equiv u_{1}/u_{0} \) is the flow precompression in
 the CR precursor.

For determining the CR diffusion coefficient \( \kappa \) one needs to
write the wave kinetic equation for which we use the eikonal
approximation

\begin{equation}
\label{wke}
\frac{\partial N_{k}}{\partial t}+\frac{\partial \omega }{\partial k}\frac{\partial N_{k}}{\partial x}-\frac{\partial \omega }{\partial x}\frac{\partial N_{k}}{\partial k}=\gamma _{k}N_{k}
\end{equation}
Here \( N_{k} \) is the number of wave quanta, \( \omega \) is the
wave frequency \( \omega =-ku+kV_{\rm A}\simeq -ku \),
\( k \) is the wave number. The left hand side has a usual Hamiltonian 
form that states the
conservation of \( N_{k} \) along the lines of constant frequency \(
\omega (k,x)=\const \) on the \( k,x \) plane. The first term on the
r.h.s. describes the wave generation on the cyclotron instability of a
slightly anisotropic particle distribution. It can be expressed
through its spatial gradient. The resonance condition \( kp=\const \)
(``resonance sharpening,''
\eg Drury \etal 1996 {[}D96{]}) is implied.

\section{Outline of the Analysis}

It is convenient to use the wave energy density normalized to \( d\ln
k \) and to the energy density of the background magnetic field \(
B^{2}_{0}/8\pi \) instead of \( N_{k} \)

\begin{equation}
\label{en:dens}
I_{k}=\frac{k^{2}V_{\rm A}}{B^{2}_{0}/8\pi }N_{k}
\end{equation}
along with the partial pressure of CRs normalized to \( d\ln p \) and
to the shock ram pressure
\( \rho _{1}u^{2}_{1} \)

\begin{equation}
P=\frac{4\pi }{3}\frac{mc^{2}}{\rho _{1}u^{2}_{1}}\frac{p^{5}}{\sqrt{p^{2}+1}}f(p,x)
\end{equation}
Using this variables, denoting \( g=P/p \), assuming a steady state
and \( p\gg 1 \), eqs.(\ref{dc1},\ref{wke}) rewrite

\begin{eqnarray}
\frac{\partial }{\partial x}\left( ug+\kappa \frac{\partial g}{\partial x}\right) =\frac{1}{3}u_{x}p\frac{\partial g}{\partial p} &  & \label{dc2} \\
u\frac{\partial I}{\partial x}+u_{x}p^{3}\frac{\partial }{\partial p}\frac{I}{p^{2}}=\frac{2u_{1}^{2}}{V_{\rm A}}\frac{\partial }{\partial x}P  &  & \label{wke2} 
\end{eqnarray}
Here the CR diffusion coefficient \( \kappa \) can be expressed
through the wave intensity by \( \kappa =\kappa _{\rm B}/I \) . The
difference between these equations and those used in \eg D96 is due to
the terms with \( u_{x}\neq 0 \). Far away from the subshock where \(
u_{x}\to 0 \), one simply obtains \(I=2u_{1}P/V_{\rm A} \). The most
important change to the acceleration process comes from the terms with
\( u_{x}\neq 0 \). Indeed, let us recall first how the equation
(\ref{dc2}) may be treated in the linear case \( u_{x}\equiv 0 \) for
\( x>0 \). Integrating both sides between some \( x>0 \) and $x=\infty $,
one obtains

\begin{equation}
\label{cd:lin}
u_{1}g+\frac{\kappa _{0}V_{\rm A}}{2u_{1}}\frac{1}{g}\frac{\partial g}{\partial x}=0
\end{equation}
where we denoted \( \kappa _{0}\equiv \kappa _{\rm B}/p\simeq const \)
for \( p\gg 1 \).  Although this equation has a formal spatial scale
\( l\sim \kappa _{0}/u_{1}M_{\rm A}g \), its only solution is a power
law in \( x \)

\begin{equation}
\label{g:bell}
g\propto 1/\left( x+x_{0}\right) 
\end{equation}
and thus has no scale. It simply states the balance between the
diffusive flux of particles escaping upstream (second term in
eq.{[}\ref{cd:lin}{]}) and their advection with thermal plasma in the
downstream direction (the first term). As we shall see, this balance
is possible not everywhere upstream and the physical reason why it
appears to be so robust in the case \( u_{x}=0 \) is that the flows of
particles and waves on the \( x,p \)-plane (including the diffusive
particle transport) are both directed along the \( x \)-axis. If,
however, the flow modification upstream is significant (\( u_{x}>0 \),
\( x>0 \)), the situation changes fundamentally.  Fig.1 explains how
the flows of particles and waves on the \( x,p \)-plane become
misaligned even though they are both advected with the thermal
plasma. In fact, the flows separate from each other and, since neither
of them can exist without the other (waves are generated by particles
that, in turn, are trapped in the shock precursor by the waves) they
both disappear in that part of the phase space where the separation is
strong enough. To understand how this happens it is useful to rewrite
eqs.(\ref{dc2}-\ref{wke2}) in the following characteristic form (we
return to the particle number density \( f \))

\begin{figure}
\vspace*{2.0mm} % just in case for shifting the figure slightly down
\includegraphics[width=8.3cm]{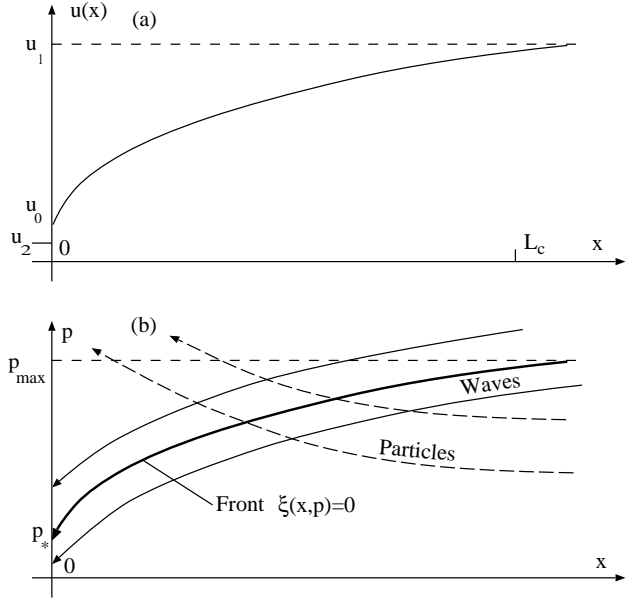}
\caption{(a) flow velocity; (b) characteristics of eqs.(\ref{dc3},\ref{wke3}).}
\end{figure}

\begin{eqnarray}
\left( u\frac{\partial }{\partial x}-\frac{1}{3}u_{x}p\frac{\partial }
{\partial p}\right) f=-\frac{\partial }{\partial x}
\kappa \frac{\partial f}{\partial x} &  & \label{dc3} \\
\left( u\frac{\partial }{\partial x}+u_{x}p\frac{\partial }
{\partial p}\right) \frac{I}{p^{2}}=\frac{2u_{1}^{2}}
{V_{\rm A}p^{2}}\frac{\partial }{\partial x}P &  & \label{wke3} 
\end{eqnarray}
One sees from the l.h.s.'s of these equations that particles are
transported towards the subshock in \( x \) and upwards in \( p \)
along the family of characteristics \( up^{3}=const \), whereas waves
move also towards the subshock but downwards in \( p \) along the
characteristics \( u/p=const \). As long as \( u(x) \) does not
significantly changes the waves and particles propagate together
(along \( x \)-axis) as \eg in the case of unmodified shock or far
away from the subshock where \( u_{x}\rightarrow 0 \). When the flow
compression becomes important (\( u_{x}\neq 0 \)) their separation
leads to decrease of both the particle and wave energy densities
towards the subshock. To describe this mathematically, let us assume
that the linear relation between \( P \) and \( I \) is still a
reasonable approximation even if \( u_{x}
\) is nonzero but small. (A more general case is considered in a
longer version of this paper).  Then, integrating eq.(\ref{dc2})
between some \( x>0 \) and \( x=\infty \), instead of (\ref{cd:lin})
we obtain

\begin{equation}
\label{cd:int}
ug+u_{1}\frac{L}{g}\frac{\partial g}{\partial x}=-\frac{1}{3}\int
_{x}^{\infty }u_{x}p\frac{\partial g}{\partial p}dx
\end{equation}
In contrast to the solution of eq.(\ref{cd:lin}) the length scale \(
L\equiv \kappa _{0}/2u_{1}M_{\rm A} \) enters the solution of this
equation. This is because it has a nonzero r.h.s. The solution of this
equation changes rapidly on a scale \( L\ll L_{\rm c} \) where \(
L_{\rm c}=\kappa (p_{\rm max})/u_{1} \) is the total scale height of
the CR precursor. In addition to \( x \) and \( p \), we introduce a
fast (internal) variable \( \xi (x,p) \) as follows

\begin{equation}
\label{ksi}
\xi =\frac{x-x_{\rm f}(p)}{L}
\end{equation}
 where \( x=x_{\rm f}(p) \) is some special curve on the \( x,p \) plane which bounds the
solution and will be specified later. We rewrite eq.(\ref{cd:int}) for \( \xi \; -{\rm fixed},\; L\to 0 \)
which leads to the solution

\begin{equation}
\label{g:intern}
g(\xi ,x,p)=\frac{S(x,p)}{w(p)+e^{-S\xi /u_{1}}}
\end{equation}
where \( w(p)=u_{\rm f}+(1/3)pdu_{\rm f}/dp \), \( u_{\rm f}(p)\equiv
u\left[ x_{\rm f}(p)\right] \), 
\newline
\( S(x,p)=(1/3)pdu_{\rm f}/dp-(1/3)\int _{x}^{\infty }u_{x}p\partial 
G/\partial pdx \) and
\newline
\( G(x,p)=\lim _{\xi \to \infty }g(\xi,x,p)=S/w \). 
Eq.(\ref{g:intern}) describes the transition in the
particle distribution between its asymptotic value \( g=G \) at \( \xi
\to \infty \) and \( g=0 \) at \( \xi \to -\infty \) as a result of
particle losses caused by the lack of resonant waves towards the
subshock. The position of the transition front (\( \xi (x,p)=0 \)) is
determined by the condition of non-secular behaviour of the next order
asymptotic expansion and must coincide with one of the characteristics
of the operator on the l.h.s. of eq.(\ref{wke3}),
\ie

\[
\left( u\frac{\partial }{\partial x}+u_{x}p\frac{\partial }{\partial p}\right) \xi (x,p)=0\]
or

\[
u_{\rm f}(p)-p\frac{du_{\rm f}}{dp}=0\]
The reasonable choice of the concrete characteristic is based on the
existence of the absolute maximum momentum \( p_{\rm max} \) beyond
which there are neither particles nor waves. That means \( u_{\rm
f}(p)\equiv u\left[ x_{\rm f}(p)\right] =u_{1}p/p_{\rm max}
\). Likewise, the function \( x=x_{\rm f}(p) \) is defined by \(
x_{\rm f}(p)=u^{-1}\left( u_{1}p/p_{\rm max}\right) \).

\paragraph{External solution\label{sec: ext}}

While having obtained the form and the position \( x=x_{\rm f}(p) \)
of the narrow front in the particle distribution \( g(x,p) \) we still
need to calculate \( g \) to the right from the front where it decays
with \( x \). This would be the external solution \( G(x,p) \)
introduced above. It is clear that

\[
\max _{x}g(x,p)\approx G(x_{\rm f},p)\equiv G_{0}(p)\]
so that from eq.(\ref{cd:int}) we have the following equation

\begin{equation}
\label{cd:ext}
u_{\rm f}(p)G_{0}(p)=-\frac{1}{3}\int _{x_{\rm f(p)}}^{\infty
}u_{x}p\frac{\partial G}{\partial p}dx
\end{equation}
The most important information about \( G(x,p) \) is contained in \(
G_{0}(p) \) for which from the last equation we obtain

\begin{equation}
\label{G0:eq}
\frac{\partial }{\partial p}v(p)G_{0}(p)+4\frac{u_{1}}{p_{1}}G_{0}(p)=0
\end{equation}
where we have introduced \( v(p) \) by

\begin{equation}
\label{v:def}
v(p)=\frac{1}{G_{0}(p)}\int ^{u_{1}}_{u_{\rm f}(p)}G(x,p)du(x)
\end{equation}
Eq.(\ref{G0:eq}) can be easily solved for \( G_{0} \)

\begin{equation}
\label{G0:sol}
G_{0}(p)=\frac{C}{v(p)}\exp \left( -4\frac{u_{1}}{p_{\max }}\int
\frac{dp}{v(p)}\right)
\end{equation}
(where \( C \) is a constant). However, the function \( v \) depends
on the solution itself.  Nevertheless, it can be calculated prior to
determining \( G_{0} \) and therefore, this solution may be written in
a closed form. In the case \( p\simeq p_{\max } \)one obtains (the
shape of the cut-off)

\begin{equation}
\label{G0:pmax}
G_{0}\sim \left( p_{\max }-p\right) ^{3}
\end{equation}
In the rest of the \( x,p \)-domain where \( x_{\rm f}(p)<x<L_{\rm c}
\) and \( p \) is not close to \( p_{\max } \), we may assume that the
CR diffusion coefficient is close to its Bohm value since in contrast
to the phase space region \( x\approx x_{\rm f}(p) \) at each given \(
x,p \) there are waves generated along the entire characteristic of
eq.(\ref{wke2}) passing through this point of the phase space and
occupying an extended region of the CR precursor, Fig.1. We may adopt
then the asymptotic high Mach number solution (Malkov and Drury 2001
{[}MD{]} and references therein)

\[
G(x,p)\simeq G_{0}(p)\exp \left( -\frac{1}{\kappa _{\rm B}}\int
^{x}_{x_{\rm f}}udx\right) \]
Using the linear approximation for \( u(x) \) {[}MD{]} \(
u=u_{0}+u_{1}x/L^{\prime }_{\rm c} \) with \( L^{\prime }_{\rm
c}=\kappa _{\rm B}(\hat{p})/u_{1} \), where it is implied that the
maximum contribution to the particle pressure comes from some momentum
\( p\sim \hat{p} \), we can express \( v \) in the form of an error function
integral

\[
v(p)\simeq \int ^{u_{1}}_{u_{\rm f}}du\exp \left[ -\frac{L^{\prime }_{\rm c}}
{2u_{1}\kappa _{\rm B}(p)}\left( u^{2}-u^{2}_{\rm f}\right) \right] \]
The further algebra simplifies in two limiting cases leading to the
following asymptotic expressions for the particle distribution

\[
G_{0}(p)=\left\{ \begin{array}{cc}
\frac{C}{\sqrt{p}}\exp \left( -8\sqrt{\frac{2p}{\pi p_{\max }}}\right) , & \frac{p_{\max }}{p}\gg 1\\
\left( p_{\max }-p\right) ^{3} & p\la p_{\max }
\end{array}\right. \]
This result is valid for \( p\ge p_{*}\equiv p_{\max }/R=p_{\max
}u_{0}/u_{1}\la \hat{p} \), whereas for \( p<p_{*} \) one still has \(
G_{0}=C/p^{q(p)-3} \), where the detailed calculations of \( q(p) \),
\( C \) and $R $ may be found in MD.

\section{Conclusions}

Refraction to shorter wave lengths in a nonlinearly modified flow
results in a spectral break at \( p=p_{*}=p_{max}/R \) where \( R=u_1/u_0\gg 1
\) is the nonlinear pre-compression of the flow. The spectrum rollover
(\( p_{max}/R<p<p_{max} \)) is described by \( f\propto
p^{-q}e^{-\sqrt{p/\overline{p}}} \)with even faster decay at \( p\sim
p_{\rm max} \) while an approximate power-law \( p^{-q(p)} \) is
still valid at \( p<p_{*} \). Since \( R \) itself is proportional to
\( p_{*} \), the spectral break \( p_{*} \) should grow slower than \(
\sqrt{p_{max}} \). Due to the lack of waves at \( p_{max} \), it
advances slower than in the Bohm case.

\begin{acknowledgements}

We acknowledge helpful discussions with R.D. Blandford, R.Z. Sagdeev
and H.J. V\"olk. This work was supported by U.S. DOE under Grant
No. FG03-88ER53275.

\end{acknowledgements}

\end{document}